# Dynamic Color Displays Using Stepwise Cavity Resonators

Yiqin Chen[1,#], Xiaoyang Duan[2,3,#], Marcus Matuschek[3,#], Yanming Zhou[1], Frank Neubrech[3], Huigao Duan[1,*], Na Liu[2,3,*]

[1]State Key Laboratory of Advanced Design and Manufacturing for Vehicle Body, College of Mechanical and Vehicle Engineering, Hunan University, 410082 Changsha, P. R. China.

[2]Max Planck Institute for Intelligent Systems, Heisenbergstrasse 3, 70569 Stuttgart, Germany.

[3]Kirchhoff Institute for Physics, University of Heidelberg, Im Neuenheimer Feld 227, 69120 Heidelberg, Germany.

[#] These authors contribute equally to this work.

[*]Corresponding author: na.liu@kip.uni-heidelberg.de and duanhg@hnu.edu.cn

Tel: 0049 711 6891838

Abstract

High-resolution multicolor printing based on pixelated optical nanostructures is of great importance for promoting advances in color display science. So far, most of the work in this field has been focused on achieving static colors, limiting many potential applications. This inevitably calls for the development of dynamic color displays with advanced and innovative functionalities. In this Letter, we demonstrate a novel dynamic color printing scheme using magnesium-based pixelated Fabry-Pérot cavities by grey-scale nanolithography. With controlled hydrogenation and dehydrogenation, magnesium undergoes unique metal and dielectric transitions, enabling distinct blank and color states from the pixelated Fabry-Pérot resonators. Following such a scheme, we first demonstrate dynamic Ishihara plates, in which the encrypted images can only be read out using hydrogen as information decoding key. We also demonstrate a new type of dynamic color generation, which enables fascinating transformations between black/white printing and color printing with fine tonal tuning. Our work will find wide-ranging applications in full-color printing and displays, colorimetric sensing, information encryption and anti-counterfeiting.

Color printing based on engineered optical nanostructures represents an important step forward in optical science, as it fosters a variety of applications for high-resolution color displays,[1-11] color filters,[12-19] high-density optical data storage,[20,21] commercial anti-counterfeiting and data encryption.[19,22,23] Compared to conventional pigment-based color generation, structural colors hold the advantages of higher pixel resolution, higher data density, higher compactness, and enhanced stability without color fading. Using modern nanofabrication techniques, pixelated nanostructures with controlled size, geometry, and arrangement can be well defined, enabling precise tuning of the structural colors pixel by pixel via selective reflection, transmission, or scattering of light. Since the first demonstration of full-color printing using plasmonic nanoparticles by Kumar *et al.*,[1] this research field has flourished with significant advances, particularly on the pursuit of advanced functionalities with lower cost and better color tunability, taking the advantage of plasmonic resonances or optical interference effects.

So far, most of the research efforts in this field have been exerted on static color printing. A transition towards dynamic color printing is inevitable and imperative, as dynamic colors carry much richer information and can enable a wealthy of advanced functionalities. In general, there are two schemes to create dynamic color displays using optical nanostructures. One is to introduce dynamic materials in the surroundings of optical structure pixels[22,24-27]. As a result, the spectral profiles of the optical nanostructures can be influenced and indirectly controlled by external stimuli including electric field, heat, light, *etc.*, thus leading to dynamic color changes. The other is to utilize optical nanostructures made of dynamic materials as pixels.[28] Such dynamic pixels can directly respond to external stimuli for achieving color changes. Nevertheless, optical materials that possess both reversible dynamic responses and distinct optical behavior in the visible spectral range are not choiceful. Magnesium (Mg) is one of the promising candidates, as it exhibits excellent optical properties at high frequencies and also can absorb/desorb hydrogen, undergoing reversible transitions between metal and dielectric hydride ($MgH_2$) states.[29,30] Very recently, dynamic color tuning has been accomplished through controlled hydrogenation/dehydrogenation of Mg nanoparticles, which served as dynamic pixels in color microprints.[28]

In this Letter, we introduce a third scheme to create dynamic color displays. Here, Mg is not only a surrounding material, but also takes part in constituting dynamic pixels. More specifically, we utilize grey-scale nanolithography to generate stepwise hydrogen silsesquioxane (HSQ) pillars, which are sandwiched between a thick aluminum (Al) film as back mirror and a metallic capping layer composed of Mg/titanium (Ti)/palladium (Pd). This results in stepwise pixelated Fabry-Pérot (FP) cavities with different cavity heights. Upon hydrogen absorption and desorption, the Mg layer can be reversibly switched between a reflective metal state and a dielectric hydride state, enabling vivid color changes from individual FP resonators. Due to the well-modulated and sharp FP resonances, the generated colors are much more brilliant and richer than those from dynamic displays composed of Mg nanoparticles. In addition, this scheme utilizes the Mg layer directly from thin film deposition to achieve dynamic color changes without any post-nanofabrication steps. This avoids exhaustive optimization and special care for catalytic nanoparticle manufacturing, which can significantly influence the catalytic efficiency and performance. Following such a scheme, we first demonstrate dynamic Ishihara plates, in which the encrypted images can only be read out using hydrogen as information decoding key. Taking a further step, we demonstrate a novel dynamic color display, which exhibits fascinating transformations between black/white printing and color printing.

The working scheme of our dynamic color printing is illustrated in Figure 1. HSQ pillars of different heights are patterned on a thick Al film (100 nm) using grey-scale nanolithography as shown in Fig. 1a. The fabrication details can be found in Supporting Information. The capping layer consists of Mg/Ti/Pd materials (50 nm/2 nm/3 nm). This metallic capping layer, stepwise HSQ pillars, and the Al mirror form a series of FP cavities. Before hydrogenation, the 55 nm Mg/Ti/Pd capping layer efficiently reflects the visible light, resulting in no color generation. This defines a blank state (see Fig. 1b). Upon hydrogen exposure, Pd catalyzes the dissociation of hydrogen molecules into hydrogen atoms, which then diffuse through Ti into the Mg layer. Ti is adopted here to prevent Mg and Pd from easy alloying.[31,32] As Mg is gradually transformed to $MgH_2$, the effective thickness of the metallic capping layer decreases and light starts to transmit through it. When Mg is fully hydrogenated into $MgH_2$, a variety of FP interference resonators are formed and colors are selectively reflected (see Fig. 1b). In this case, each FP resonator consists of a $TiH_2$/PdH capping layer, a double dielectric spacer ($MgH_2$ + HSQ), and an Al back mirror. Such asymmetrical FP resonators with ultrathin lossy capping can generate vivid and high-contrast colors with a wide gamut,[33,34] representing a color state. The resonance properties such as the reflectance peak positions and the number of allowed modes in the FP resonators are largely governed by the individual cavity heights (see Fig. 1b).

Crucially for applications, the color state can be switched back to the blank state using oxygen as shown in Fig. 1a. The oxidative dehydrogenation process involves binding of oxygen with the desorbed hydrogen atoms from $MgH_2$ to form $H_2O$.[35,36] This avoids a buildup of hydrogen at the Pd surface, thus facilitating hydrogen desorption. When $MgH_2$ is fully dehydrogenated into Mg, the visible light is efficiently reflected by the thick metallic capping layer (Mg/Ti/Pd) and the palette reaches the blank state again. It is noteworthy that the presented scheme here is in evident difference from our previous work using Mg nanoparticles as dynamic pixels.[28] In the latter case, the color state was always the initial state and hydrogen was used majorly to erase images or information, whereas in our new scheme the blank state is the initial state and hydrogen is employed to uncover images or information. In this regard, the new scheme offers a much higher level of information encryption.

In order to render a palette of colors with a broad range, successive resonance tuning of the FP cavities is carried out by varying the electron exposure dose ($D$) and filling factor ($F$). Each square-shaped tile in the palette has a side length of 20 μm and contains 40 × 40 HSQ pillars arranged in a lattice with a fixed periodicity ($P$) of 500 nm along both directions. $F$ is defined as the ratio between the electron exposure area and the lattice area ($P^2$). Fig. 2a shows an optical microscopy image of the palette in the blank state captured using a 10× objective with a numerical aperture (NA) of 0.3 in a reflection bright-field mode. Along the horizontal axis, the electron exposure dose increases from 29 to 118 μC/cm², while along the vertical axis the filling factor decreases from 94% to 38%. Enlarging $D$ leads to the increase of the pillar height, while enlarging $F$ increases the HSQ pillar area size and also slightly increases the pillar height. Altogether, this gives rise to stepwise pillar height tuning from 250 nm to 460 nm in the palette. As shown in Fig. 2a, overall the palette does not exhibit colors in the blank state. The slightly dull colors in the upper regions of the palette is mainly attributed to the suppressed reflectance from the tiles with small filling factors due to strong scattering from the HSQ pillars capped with rough metals (see Supporting Information). In addition, for tiles with small filling factors, the deposited metals can be directly on the Al mirror. The resulting dull colors can be elegantly utilized for advanced display applications, which will be discussed in Fig. 5. The optical microscopy image of the palette after hydrogenation in the color state is presented in Fig. 2b, in which brilliant colors covering a wide gamut are observed. The blank and color states can be reversibly switched through controlled hydrogenation and dehydrogenation. In this work, 0.2% hydrogen and 20% oxygen at 80°C were employed for facilitating complete hydrogenation and dehydrogenation of the Mg films, respectively. A control experiment to validate the crucial role of Mg for dynamic color displays was also carried out (see Supporting Information). In the absence of Mg, a palette composed of a Ti/Pd (2 nm/3 nm) capping layer, stepwise HSQ pillars, and an Al mirror does not show dynamic color switching through hydrogenation/dehydrogenation.

To further understand the dynamic color generation, the spectral evolutions of the representative tiles during hydrogenation are presented in Fig. 3a. These tiles (i-iv) with increasing cavity heights are highlighted in Figs. 2a and 2b in the blank and color states, respectively. Before hydrogenation, these tiles appear 'white'. This is also confirmed by their overall broad spectral profiles of the measured reflectance spectra in the visible range. We first examine the case in Fig. 3a (i). Within the broad reflectance spectrum, there is a tiny kink observable near 505 nm. This corresponds to the FP resonance arising from the cavity formed between the Mg/Ti/Pd capping layer and the Al mirror. Upon hydrogen exposure, this FP resonance becomes more and more prominent over time, eventually manifesting itself as a distinct reflectance peak. It is also accompanied with successive intensity decreases of the two resonance shoulders as highlighted by the grey dashed lines in Fig. 3a (i). Simultaneously, the reflectance peak exhibits successive red-shifts, until it halts at approximately 584 nm. Such red-shifts are due to the transition of Mg into $MgH_2$ through hydrogenation, which successively leads to a final cavity with double spacer, HSQ+$MgH_2$. This results in a pronounced FP resonance, giving rise to a yellowish color. Such drastic modifications of the resonance behavior through hydrogenation render sharp transformations between the blank and color states possible. As the HSQ pillar height increases (see from Figs. 3a (i) to (iv)), the reflectance peak shifts to longer wavelengths and meanwhile a new reflectance peak emerges at shorter wavelengths. This corresponds to a higher order FP resonance, which further helps to alter the spectral profile, enabling color generation with a wide gamut. Notably, the colors generated from this Mg-based FP resonator scheme are more vivid and show higher contrast than those from our previous Mg particle scheme,[28] in that the new scheme offers much sharper and pronounced resonances. In addition, the possibility to largely tune the resonance positions and greatly modify the spectral profiles by dynamically changing the effective cavity heights through hydrogenation of a simple Mg layer suggests a straightforward solution to achieving dynamic color displays for practical applications.

To provide a deeper insight into the resonance behavior, numerical simulations were carried out for the configuration in Fig. 3a (iii), as it nicely involves two FP resonances in the color state. In the simulation, the height of the HSQ pillar was taken as 440 nm. As shown in Fig. 3b, the simulated reflectance spectra in the blank state (grey line) and color state (black line) overall agree well with the measured data. The relatively low reflectance at shorter wavelengths is due to strong scattering at rough metal surfaces (see Supporting Information). The electric field distributions at 649 nm (grey solid arrow) and 432 nm (grey hollow arrow) in the blank state reveal the formation of the $2^{nd}$ and $3^{rd}$ order FP resonances formed between the Mg/Ti/Pd capping layer and the Al mirror, respectively. In the color state, the electric field distributions at 715 nm (blue solid arrow) and 485 nm (blue hollow arrow) suggest the formation of the $2^{nd}$ and $3^{rd}$ orders of the FP cavity modes formed between the TiH$_2$/PdH layer and Al mirror, respectively. The FP resonances in the color state are much more pronounced than those in the blank state due to the ultrathin capping layer after hydrogenation, which allows for more light to pass through and therefore enables much stronger interference effects.

The drastic color transformations between the blank and color states constitute an ideal platform for information encryption in microprints. As a demonstration, we have designed and fabricated two dynamic Ishihara color plates using the colors selected from the palette. Different information is encoded in the two plates. Ishihara plates are often used in clinics for color vision tests. Generally, a symbol is made with colored dots placed in a background comprising dots of a different color. Figs. 4a (i) and 4b (i) present the optical microscopy images of the two Ishihara plates fabricated following the same procedures for the palette. We deliberately constructed the two Ishihara plates using only the 'white' pixels selected from the palette so that the plates display no information before hydrogenation. The enlarged scanning electron microscopy (SEM) images of the two plates are also shown in Fig. 4. Each plate contains many dotted areas. In each dotted area, HSQ pillars with lattice spacing of 500 nm are sandwiched between a 55 nm Mg/Ti/Pd capping layer and an Al mirror. In different dotted areas, the HSQ pillar heights can be readily tuned by grey-scale nanolithography to generate designated colors. For the unpatterned regions in between the dotted areas, the Mg/Ti/Pd layer directly resides on the Al mirror.

The information encrypted in the two Ishihara plates can be decoded after hydrogenation as shown in Figs. 4a (ii) and 4b (ii), respectively. In Fig. 4a (ii), '27' in magenta is revealed against a gradually changing bluish-greenish color background with high contrast. Alternatively, in Fig. 4b (ii), a 'cat' symbol in orange is displayed against a yellowish background. Although the two hues in Fig. 4b (ii) have a small discrepancy in the color space, the plate presents a high-contrast image, demonstrating precise generation of the designated colors. The unpatterned areas in both plates exhibit a uniform black background color, further enhancing the contrast of the Ishihara color plates. The spectral evolution of the background during hydrogenation can be found in Supporting Information, in which broadband low reflectance, *i.e.*, high absorbance gradually comes into existence. These experimental results elucidate that our scheme is ideally suited for highly secure information encryption and anti-counterfeiting applications.

To demonstrate the possibility of creating arbitrary dynamic microprints with excellent color and tonal control, we have fabricated a color display using '*The Starry Night*' from Vincent van Gogh as design blueprint. A database was first created using the palette colors in Fig. 2b. This database includes a library of Red-Green-Blue (RGB) values as well as their corresponding electron exposure doses and filling factors. The computer-generated layout as shown in Fig. 5a (i) was obtained using a MATLAB script through color matching. The SEM image of the fabricated display is included in Fig. 5a (ii). The performance of the dynamic color display presented using snapshot images is shown in Fig. 5b and the accompanying video can be found in Supporting Movie 1. We particularly utilized the dull colors in the palette to create a black/white display for the initial state (see Fig. 5b, at $t = 0$ s), in which only the tree, mountains, and houses are clearly visible. Upon hydrogen exposure, abrupt color alterations take place. The sky and the mountains are dynamically tinted with brilliant bluish and yellowish colors. At $t = 78$ s, the display reaches the final color state, beautifully reproducing the details of the blueprint. Upon oxygen exposure, the display can be gradually restored to the initial black/white state within 35 s. To demonstrate the good durability of the transformations between the black/white and color printing, operation of the dynamic display in a number of cycles is shown in Supporting Movie 1. The durability can be further enhanced by alloying Mg with Yttrium to reduce hysteresis.[37]

In conclusion, we have demonstrated a novel scheme for generating dynamic color displays using Mg-based pixelated FP cavities. The unique hydrogenation/dehydrogenation kinetics of Mg enables dynamic alterations to the FP resonances, resulting in drastic color changes. This scheme provides a promising solution to achieving dynamic color displays with high contrast, wide gamut, and advanced properties including high-level information encryption and fascinating transformations between black/white and color microprints. For large-area manufacturing at low cost, one can use roll-to-roll or roll-to-plate nanoimprint. In addition, polytetrafluoroethylene protective coating can be applied to avoid water staining on the display surface.[38,39] Our work will shed light on creation of anti-counterfeiting and high-resolution chromatic devices with unprecedented functionalities.

# ACKNOWLEDGEMENT

This project was supported by the Sofja Kovalevskaja grant from the Alexander von Humboldt-Foundation and the European Research Council (*ERC Dynamic Nano*) grant as well as the National Natural Science Foundation of China (Grants 11574078), the Foundation for the authors of National Excellent Doctoral Dissertation of China (201318), and the Natural Science Foundation of Hunan Province (2015JJ1008, 2015RS4024). We are grateful for the valuable discussions with R. Griessen. The authors thank the support by the Max-Planck Institute for Solid State Research for the usage of clean room facilities and also thank the 4$^{th}$ Physics Institute at the University of Stuttgart for the usage of their electron-gun evaporation system. N.L. and H.D. conceived the project. Y.C. and Y.Z. performed the grey-scale electron-beam lithography. X.D. and M.M. performed the optical measurements. X.D. carried out the theoretical calculations. M.M. did the Mg material processing. F.N. commented on the results. All authors analyzed the data and wrote the manuscript.

# ASSOCIATED CONTENT

Supporting Information

The Supporting Information is available free of charge on the ACS Publications website at DOI:

Details on sample fabrication, optical measurements, dull colors in the palette, CIE chromaticity diagram analysis, control experiment without Mg, numerical simulations, black background in the Ishihara plate, and viewing angle dependence (PDF)

# AUTHOR INFORMATION

Corresponding Authors
*E-mails: na.liu@kip.uni-heidelberg.de and duanhg@hnu.edu.cn
Notes
The authors declare no competing financial interest.

For TOC only

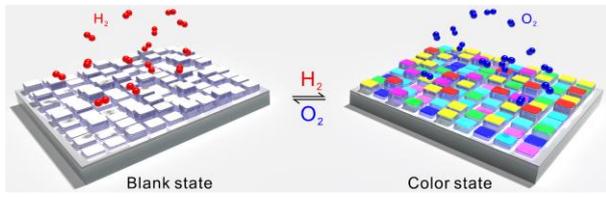

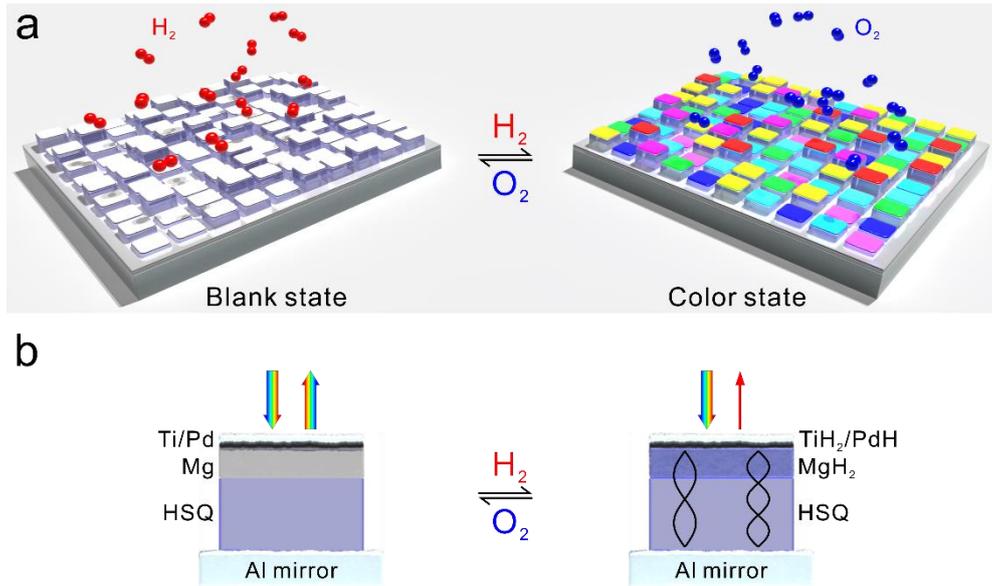

**Figure 1**. (a) Schematic of the dynamic color display using stepwise FP resonators. Pixelated HSQ pillars of different heights generated by grey-scale nanolithography are sandwiched between a Mg/Ti/Pd (50 nm/2 nm/3 nm) capping layer and an Al mirror. Before hydrogenation, the palette is in the blank state. After hydrogenation, the palette reaches the color state through the transition of Mg to $MgH_2$, which allows for the formation of a series of FP resonators with different cavity heights. The colors can be erased using oxygen through transition of $MgH_2$ back to Mg. (b) Before hydrogenation, due to the Mg/Ti/Pd capping layer, the visible light is efficiently reflected. After hydrogenation, the ultrathin capping layer $TiH_2/PdH$ allows light to pass through the double dieletric spacer ($MgH_2$+HSQ) and reaches the Al mirror. FP modes of different orders can be formed in the cavity, selectively reflecting light with specific colors.

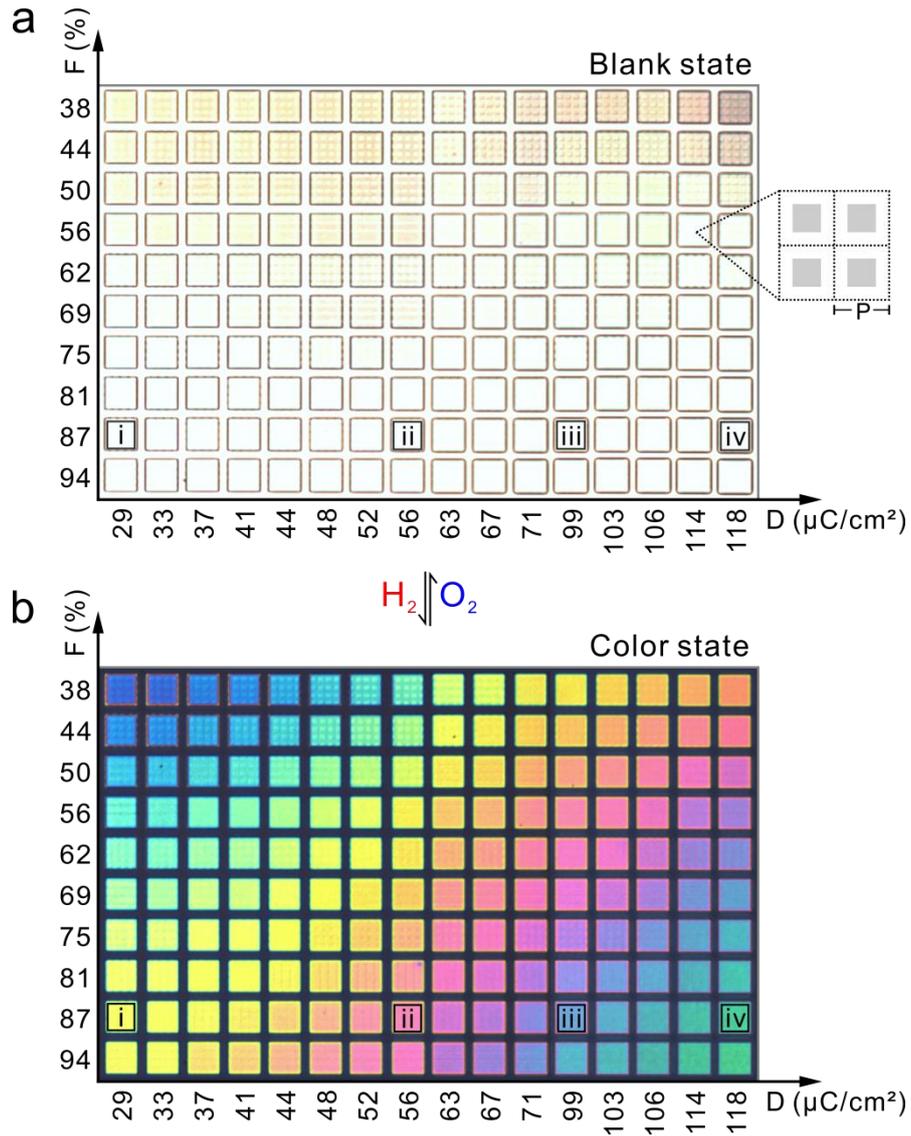

**Figure 2**. (a) Optical microscopy image of the palette in the blank state. Successive resonance tuning of the FP cavities is carried out by varying the electron exposure dose ($D$) along the horizontal axis and filling factor ($F$) along the vertical axis. Each tile in the palette has a side size of 20 μm. It contains 40 × 40 HSQ pillars arranged in a lattice with a fixed periodicity ($P$) of 500 nm along both directions. $F$ is defined as the ratio between the electron exposure area and the lattice area ($P^2$). (b) Optical microscopy image of the palette in the color state. Selective tiles are highlighted using black frames for spectral analysis in Fig. 3.

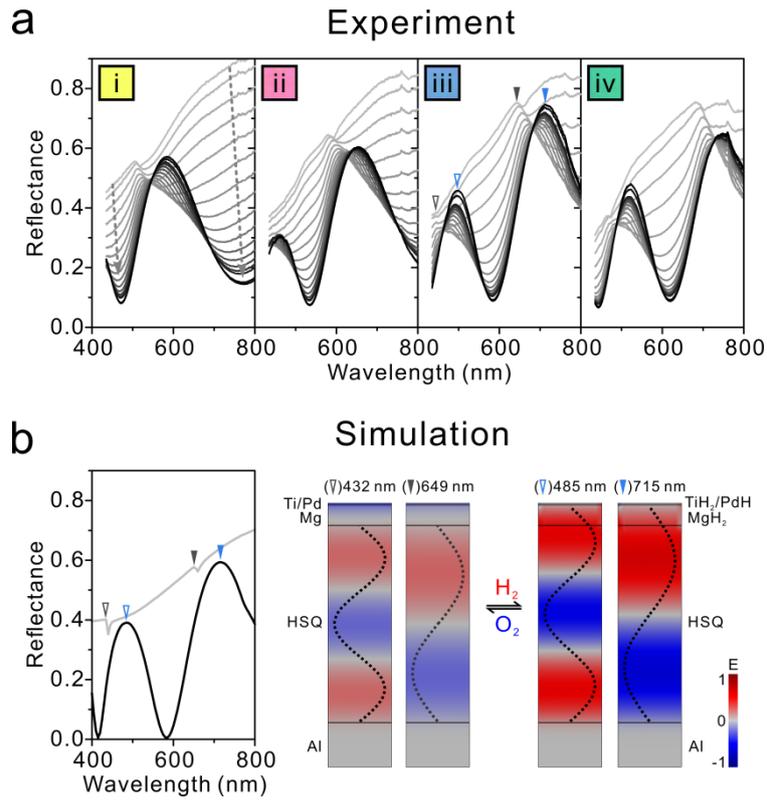

**Figure 3**. (a) Experimental spectral evolution of the representative tiles highlighted in Fig. 2 during hydrogenation. The reflectance spectra are developed from overall broad profiles (grey) to sharp FP resonances (black). (b) Simulated reflectance spectra of tile (iii) in (a) in the blank (grey line) and color (black line) states, respectively. In the simulation, the height of the HSQ pillar was taken as 440 nm. The electric field distributions of the different FP resonances (highlighted using arrows) before and after hydrogenation.

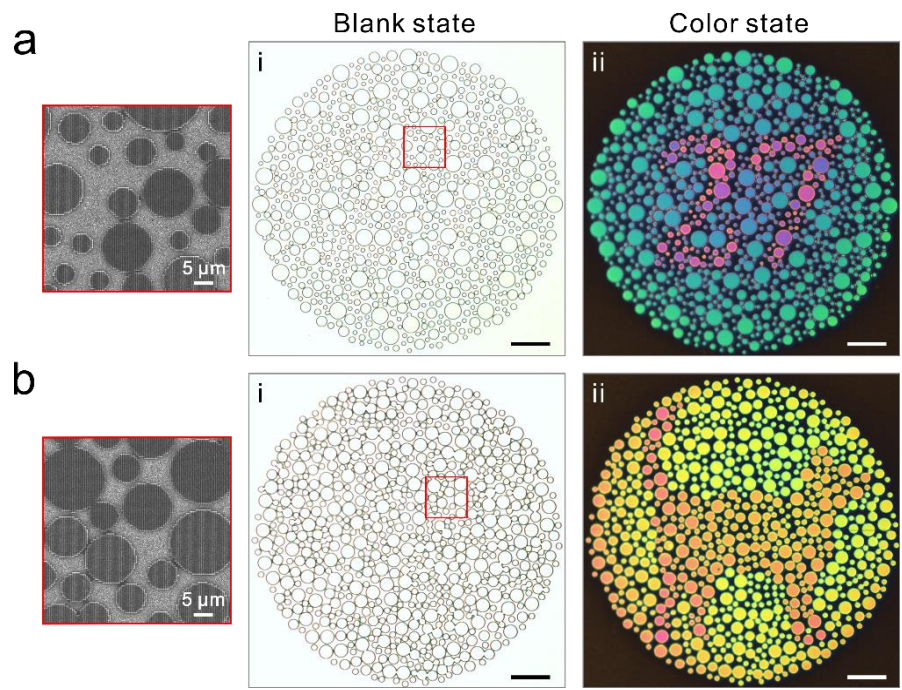

**Figure 4**. Optical microscopy images of the Ishihara display '27' (a) and 'cat' display (b) in the blank (i) and color (ii) states. Enlarged SEM images of the highlighted areas. Scale bar: 50 μm.

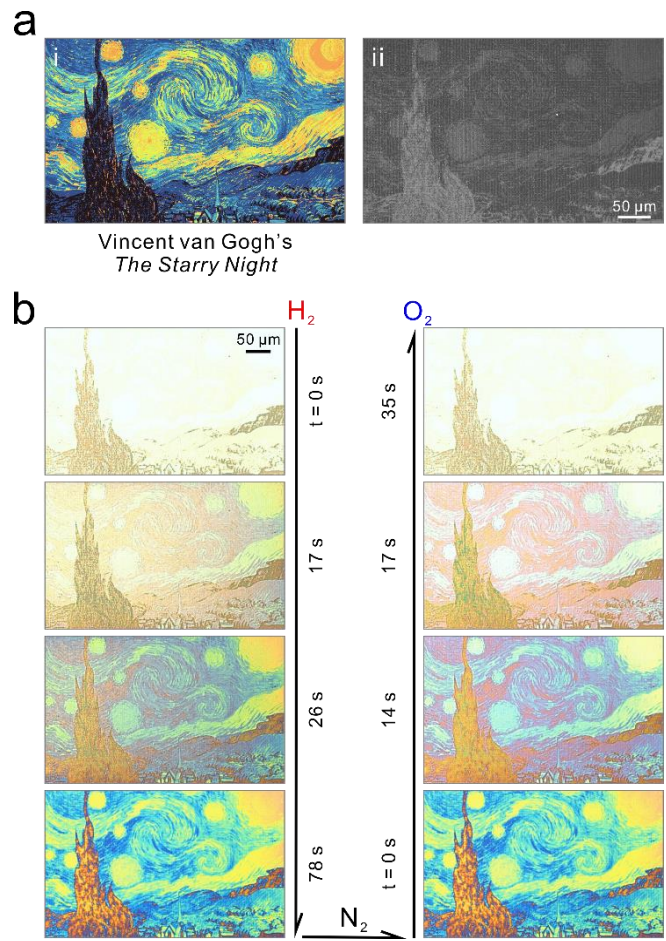

**Figure 5**. (a) Computer-generated layout of Vincent van Gogh's *The Starry Night* using the color database obtained from the palette and the overview SEM image of the fabricated display. (b) Dynamic display of the artwork, showing transformations between black/white printing and color printing.